\documentstyle[11pt,aasms4]{article}

\def\eg{{\it e.g.}}
\def\ie{{\it i.e.~}}

\def \mdot {\rm{M}$_\odot$~yr$^{-1}$}
\def \kms{km~$\rm{s}^{-1}$}

\begin{document}

\title{\LARGE \bf Where's the Doughnut?\\
                  LBV bubbles and Aspherical Fast Winds\altaffilmark{8}}

\author{\Large Adam Frank \altaffilmark{1,5},
               Dongsu Ryu \altaffilmark{2,3,6},
           and Kris Davidson \altaffilmark{4,7}}

\altaffiltext{1}{Dept. of Physics and Astronomy, Bausch and Lomb Building,
Univ. of Rochester, Rochester, NY 14627-0171}
\altaffiltext{2}{Department of Astronomy, University of Washington,
Box 351580, Seattle, WA 98195-1580}
\altaffiltext{3}{Department of Astronomy \& Space Science, Chungnam National
University, Daejeon 305-764, Korea}
\altaffiltext{4}{Department of Astronomy, University of Minnesota,
Minneapolis, MN 55455}
\altaffiltext{5}{email: afrank@alethea.pas.rochester.edu}
\altaffiltext{6}{email: ryu@hermes.astro.washington.edu}
\altaffiltext{7}{email: kd@ast1.spa.umn.edu}
\altaffiltext{8}
{Submitted to the Astrophysical Journal}

\begin{abstract} 
In this paper we address the issue of the origin of
LBV bipolar bubbles. Previous studies have explained the
shapes of LBV nebulae, such as $\eta$ Car, by invoking the interaction
of an isotropic fast wind with a previously deposited, slow
aspherical wind (a ``slow torus").  In this paper we focus on the
opposite scenario where an {\it aspherical fast wind} expands into a
previously deposited {\it isotropic slow wind}. Using high resolution
hydrodynamic simulations, which include the effects of radiative cooling,
we have completed a series of numerical experiments to test if and how
aspherical fast winds effect wind blown bubble morphologies.  Our
experiments explore a variety of models for the latitudinal variations
of fast wind flow parameters. The simulations demonstrate that
aspherical fast winds can produce strongly bipolar outflows.  In
addition the properties of outflows recover some important aspects of
LBV bubbles which the previous "slow torus" models can
not.  
\end{abstract}


\clearpage

\section{Introduction}

In just a few years the HST has transformed our
understanding of the massive unstable stars know as Luminous Blue
Variables (LBVs).  Recent observations have revealed a number of LBVs
or LBV candidates to be surrounded by extended {\it aspherical}
outflows. The most extraordinary of these is the markedly bipolar
nebula surrounding $\eta$ Carinae (``the homunculus'': 
\cite{Hesterea91}; \cite{Ebbetsea93}; \cite{HumDav94}).  Other LBVs show
nebulae with varying degrees of asphericity from elliptical
(R127: \cite{Notaea95}) to strongly bipolar (which we define though the
presence of an equatorial waist: HR Carinae: \cite{Notaea95};
\cite{Weisea96}).

These morphologies are quite similar to what has been observed in
Planetary Nebulae (PNe) which arise from low mass stars
(\cite{Balick87}; \cite{Machadoea96}). The aspherical shapes of PNe
have been successfully explained through a scenario termed the ``Generalized
Interacting Stellar Winds'' model (GISW:
\cite{Kwok78}, \cite{Kwok82}, \cite{Kahn83}, \cite{Frankea93}; \cite{FM94}; 
\cite{MF95}).  In the GISW model an
isotropic fast wind from the central star (a proto-white dwarf) expands
into an aspherical (toroidal) slow wind ejected by the star in its
previous incarnation as a Asymptotic Red Giant. High densities in the
equatorial plane constrain the expansion of the fast wind.  The
expanding shock which forms quickly assumes an
elliptical prolate geometry.  If the ratio of mass density between
the equator and pole (a parameter we call $q$, $q_\rho = \rho_e/\rho_p$) is 
high enough, then the elliptical bubble eventually develops a 
waist and becomes bipolar.

The similarity of PNe and LBV nebulae has led to the suggestion that
both families of objects are shaped in similar ways.  
In \cite{FBD95} (hereafter: FBD) a
GISW model for $\eta$ Car was presented in which a spherical ``outburst''
wind expelled during the 1840 outburst expanded into a toroidal
``pre-outburst'' wind.  FBD showed that the resulting bipolar outflow 
could recover both the gross morphology and kinematics of the Homunculus.
\cite{Notaea95} (hereafter NLCS) used a similar model for other
LBV nebulae presenting a unified picture of the development of LBV outflows.
More recently \cite{MacLowea96} (hereafter GLM) presented a model which
also relied on the GISW scenario but which changed the order of importance
of the winds.  The novel aspect of GLM's study was to include the 
effects of stellar rotation.  Using the Wind Compressed Disk model
of \cite{BjCas92}, GLM showed that a strong equator to pole density contrast
would likely form {\it during the outburst} when the star is close to the
Eddington limit and rotation can deflect wind
streamlines toward the equator.  In their model it is the
``post-outburst" mass loss (which was not considered in either 
FBD or NLCS) that acts as the fast wind.  The post-outburst wind in
GLM's model ``inflates'' the bipolar bubble via its interaction with the
toroidal outburst wind.

All these models have demonstrated the potential efficacy of the
GISW scenario by recovering the basic shapes observed in LBV nebulae.
However, by relying on the presence of a slow torus
they are are troubling in their mutual inconsistency.  Specifically
the question ``Where is the torus?'' must be answered.  Does the
torus form during the outburst phase as in GLM or does it form in the 
pre-outburst wind as in FBD and NLCS?  Without invoking binary interactions
or a pre-existing disk left over from the stellar formation process,
it may be difficult to get a strong toroidal density contrast in the 
pre-outburst environment.  

Stepping back further one can also ask if a disk is needed at all?  The
latter question arises from consideration of new HST images of $\eta$
Car (\cite{Morseea97}) which reveal the disk to be so highly fragmented
that it may be more reasonable to consider the structure to be a ``skirt''
of individual clumps of ejecta rather than a continuous feature.  This
point is crucial since a discontinuous equatorial spray of 
isolated bullets can not
hydrodynamically constrain an isotropic stellar wind to form a bipolar
outflow.  Thus we are led to consider an alternative model
to the one proposed by FBD, NLCS and GLM which further generalizes the
GISW model by turning that scenario on its head.  It what follows we
consider the case of an {\it aspherical fast wind} interacting with an
{\it isotropic slow wind}.  We imagine a fast wind ejected with higher
velocity along the poles than along the equator.  The question we wish
to answer is can such a wind, expanding into an isotropic environment,
account for the shapes of LBV nebulae.

There are a number of reasons for pursuing this line of investigation
some of which were cited above as criticisms of the ``classic'' GISW
model.  More importantly, however, theoretical models admit the
possibility of aspherical fast winds in massive stars.
\cite{PaulPuls90} have shown that a discontinuity (bistability)
in mass loss and
velocity occurs when the effective gravity of the star drops below a
critical value.  \cite{LamPaul91} used these results to demonstrate
that stellar rotation can induce latitudinal changes in $g_{eff}$ 
and the optical
depth of the wind.  The change in optical depth puts the polar and
equatorial regions of the star on different sides of the bistabilty
limit.  A high velocity, low density wind forms at the poles, and low
velocity, high density wind forms along the equator.   It should be
noted that the Wind Compressed Disk (WCD) model of \cite{BjCas92} also
produces aspherical winds since the equatorial focusing occurs close to
the star.  Thus a wind that has been shaped by the WCD mechanism, if it
is expanding into a slower moving environment, should be considered an
aspherical fast wind, \ie the issue is always the velocity (and
density) of previously ejected material.  It is worth noting however
that recent numerical models of the WCD mechanism (\cite{Owockiea96})
which include non-radial line forces found inhibition of the wind
compression and mass loss in the equator.  Instead a net flow in the
pole-ward direction was formed.  This is, therefore, yet another way by
which fast winds might become aspherical. Finally, and most
importantly, there is direct evidence for asphericity in fast winds.
Observations of the wind of AG Carinae (\cite{Leithererea94}) imply a
pattern of densities and velocities from pole to equator much like that
described in \cite{LamPaul91}.  Finally we note that it is worthwhile
to pursue this kind of investigation simply because it has not
been done before.  The GISW model and its variations has been very
successful in accounting for a variety of bipolar outflow phenomena
(\cite{Melal91}; \cite{BloLund92}; \cite{FrBaLi96}).  Since the effect
of aspherical fast winds has yet to be investigated the potential of
finding useful results is high which argues for a detailed study.

We note that this paper represents an initial exploratory study.  We
are using an admittedly ad-hoc formalism to control the asphericity of the
fast wind and we have not tuned our parameters to the accepted values
for any particular LBV.  Our purpose in this paper is to map out the
broad consequences of including aspherical fast winds into the GISW
formalism with an application to LBVs as a class of outflows.
In future papers we will attempt to apply our results to individual LBVs
in an attempt to make detailed contact with observational results.

The organization of the paper is as follows: In section II we describe
the numerical method and initial conditions used in our
simulations.  In Section III we present and discuss the results of our
simulations. In section IV we present our conclusions along with a
discussion of some issues raised by the simulations.

\section{Computational Methods and Initial Conditions}

The gasdynamic interactions we wish to study are governed by the Euler
equations with a `sink' term in the energy equation due to radiation
losses.  
\begin{equation}
{\partial\rho \over \partial t} + 
{\bf \nabla} \cdot \rho {\bf u} = 0, 
\label{masscon}
\end{equation}
\begin{equation}
{\partial\rho {\bf u} \over \partial t} + 
{\bf \nabla} \cdot \rho {\bf uu} = 0 ,
\label{momcon}
\end{equation}
\begin{equation}
{\partial E \over \partial t} + 
{\bf \nabla} \cdot {\bf u} (E + p) = - {\rho \over \bar m}^2 \Lambda(T), 
\label{encon}
\end{equation}
where
\begin{equation}
E = {1 \over 2} \rho \vert {\bf u} \vert^2 + {1 \over (\gamma - 1)}p, 
\label{endef}
\end{equation}
\noindent and
\begin{equation}
p = \rho {k \over \bar m} T 
\label{pdef}
\end{equation}
\noindent In the above equations ${\bar m}$ is the mean mass per
particle
and $\gamma$ is the ratio of specific heats
which we take to be $\gamma = 5/3$.

We have carried out our study using the Total Variation Diminishing
(TVD) method of \cite{harten83} as implemented by \cite{Ryuet95}.  The TVD
code is an explicit method for solving the Euler equations
which achieves second order accuracy by finding
approximate solutions to the Riemann problem at grid boundaries
while remaining non-oscillatory at shocks through the application of a lower
order monotone scheme.
For this study we used the code configured in cylindrical coordinates
($r,z$).  

The cooling term $\Lambda(T)$ was calculated via look-up tables for
$\Lambda(T)$ taken from the coronal cooling curve of \cite{DalMc72}. 
In the TVD code the cooling is applied in between hydro timesteps
via an integration of the thermal energy($E_t$) equation,
\begin{equation}
{d E_t\over dt} = - {\dot E_t} = - {\rho \over \bar m}^2 \Lambda(T).
\label{ethdot}
\end{equation}

If the cooling time scale becomes smaller than the dynamical time scale,
the cooling term becomes a "stiff" source term.
There are several ways to the handle stiff source terms
(see, \eg, \cite{LEV97}).
First, the cooling time scale as well as the Courant condition
is considered to determine the timestep.  Then, in some situations,
the timestep is governed by the cooing time scale and becomes
uncomfortably too small.
Second,  Strang's operator splitting is employed, where the
cooling is computed by multiple steps with it own timestep during
one hydrodynamic timestep.
Third, the solution takes the form
\begin{equation}
{E_t}^{n+1} =  {E_t}^{n}\exp\left(-{{\dot E_t}^{n} \over {E_t}^{n}}\Delta t
\right)_, 
\label{etsol}
\end{equation}
where the superscript $n$ refers to the time index. 
In this method, the cooling is computed in a single step.  It produces
good results even though the cooling time scale is smaller than the 
dynamical time scale. When cooling is small, the solution 
approaches the conventional
form
\begin{equation}
{E_t}^{n+1} = {E_t}^{n} - {{\dot E_t}}^{n}\Delta t.
\label{etsol2}
\end{equation}

The third method has been used in our simulations.  The term in the
exponential can be expressed as $\Delta t / \Delta t_{c}$ where $\Delta
t_{c}$ is the local cooling timestep of the gas $\Delta t_{c} = {\dot
E_t}^{n} / E_t^{n}$.  Although the method works well for small cooling
timestep, for safety  we have used \begin{equation} \Delta t =
\min\lbrack\Delta t_{h},1.5 \Delta t_{c}\rbrack, \label{timestep}
\end{equation} where $\Delta t_{h}$ is the dynamical timestep set by
the Courant Condition.  

Care must be taken in the application of cooling near contact
discontinuities (CDs) where in typical wind blown bubble temperatures
and densities can vary by more than an order of magnitude going from
low  to high values of density and high to low values of temperature.
In simulations with less than perfect resolution the CD will be smeared
out across a number of cells.  In these intermediate zones which have
both high density and high temperature the cooling term can be
anomalously large.  This leads to the removal of
 prodigious amounts of energy from the system
at just a few zones.  To avoid these problems we have tracked the CD by
following the advection of a passive tracer $\psi$ via Eq.~(\ref{masscon}),
\ie  
\begin{equation} {\partial\psi \over \partial t} +
{\bf \nabla} \cdot \psi {\bf u} = 0. \label{phicon} 
\end{equation} 
This equation has been solved with the TVD code.  Using this tracer we can
distinguish between shocks and CDs. If the fast stellar wind has $\psi =
1$ and the slow material initially filling the grid  has $\psi = 0$
then a CD can always be marked as a location with $0.1 < \psi < 0.9$.
When a CD is encountered we replace the anomalous cooling value with
one obtained from an average of the 8 nearest neighbor cells which do
not contain the CD.  Tests of the code show that this formalism allows
us to reproduce analytically derived shock speeds (\cite{KoMc92}) for
spherical radiative bubbles in a variety of power-law, $\rho \propto
r^k$, environments.

Our numerical experiments are designed to study the evolution of
wind-blown bubbles driven by aspherical fast winds.  The environment
is always assumed to be characteristic of a previously deposited
spherically symmetric ``pre-outburst'' wind which we denote as
wind 0 with mass loss rate $\dot M_0$
and velocity $V_0$.  Thus $\rho_0 = \dot M_0/4 \pi r^2 V_0$.  For
the driving ``fast'' or ``outburst'' wind, which we denote as
wind 1, we need a formalism for setting the latitudinal ($\theta$) variations
in the wind properties, \ie $\dot M_1 = \dot M_1 (\theta)$
and $V_1 = V_1(\theta)$.
We note that since we wish to drive prolate bipolar bubbles
we always assume that the velocity at the poles is larger that at
the equator

We have chosen to explore 3 models for the pole to equator variation
in wind parameters.  Each model is based on the assumption of
a different quantity remaining constant across the face of the star.
Our three models are:  
\begin{enumerate}
\item  Constant Momentum Input 
\begin{itemize}
\item $\dot \Pi = \dot M_1 V_1$ = $Const$
\end{itemize}
\item  Constant Energy   Input 
\begin{itemize}
\item $\dot E = {1 \over 2} \dot M_1 {V_1}^2$ = $Const$
\end{itemize}
\item  Constant density 
\begin{itemize}
\item $\rho_1$ = $Const$
\end{itemize}
\end{enumerate}

If we choose our fiducial values for the density and velocity at the
equator $(\rho_{1e},V_{1e})$, then the variation of these quantities 
can be expressed as $\rho_1(\theta) = \rho_{1e} g(\theta)$, and 
$V_1(\theta) = V_{1e} f(\theta)$.
We can then write $g(\theta) \propto f(\theta)^n$. For case 1
where $\dot \Pi = Const = \dot \Pi_e$, we have
\begin{equation}
\dot \Pi = \dot M_1(\theta) V_1(\theta) =
 (4 \pi {R_*}^2 \rho_{1}(\theta) V_{1}(\theta)) V_{1}(\theta) =
(4 \pi {R_*}^2 \rho_{1e} V_{1e}) V_{1e}.
\label{case1a}
\end{equation}
Thus
\begin{equation}
g(\theta) = {1 \over f(\theta)^2}.
\label{case1b}
\end{equation}
For case 1  $n=-2$. Similarly, for cases 2 and 3 one finds $n= -3$ and $0$ 
respectively.

For $f(\theta)$ we have chosen an ad hoc function which produces a smooth 
variation in $V_1$ and $\rho_1$ from equator to pole. Written in terms of 
velocity we have
\begin{equation}
V_1(\theta) = V_{1e} \left\lbrack 1 - A { \exp{(-2 B \cos(\theta)^2}) - 1 
\over \exp{(-2 B) - 1} }  \right\rbrack^{-1},
\label{vfdef}
\end{equation}
where constant $B$ determines the shape of the equator to pole variability.
The constant A is related to the magnitude of the contrast 
$A = 1 - q_v$.  Note in what follows we shall write
$q_x = X_e/X_p$ and $X$ is either the velocity or the density.
Note that $V_p = V_e/(1 - A)$ so $V_p = V_e/q_v$.  
We take $A < 1$, so that together with Eq.~(\ref{case1b})
we have $q_v = V_e/V_p < 1$ and 
$q_\rho = \rho_e/\rho_p > 1$.  These relations express our 
assumptions about the 
velocity variation across the face of the star and its consequences for the 
density distribution given our different models.
In Fig.~1 we show a plot of the fast wind $V(\theta)$ and 
$\rho(\theta)$ for each of
our three cases.

Using the formalism described above we have carried out three sets of
numerical experiments varying the assumptions about the mass loss rates
in the successive winds between each set.  Within the first two
sets we performed
a triplet of simulations with the $\theta$ dependence of $\dot M_1$ and $V_1$
corresponding to the three cases discussed above.  In set 1 we examined
the interaction between two winds of with equal mass loss rates. These
simulations are performed to give us a baseline on the gas-dynamic flow
pattern.  In the second set of experiments we examined the interaction
between two winds where the previously ejected ``slow'' pre-outburst
wind ($\dot M_0$) had a lower mass loss rate than the outburst
wind.  We examine this case because stellar evolution models show the
mass loss rate increases during an LBV eruption (\cite{Langerea94}). In
the third set of experiments
 we examined the interaction of {\it three} winds where $\dot M_0 = \dot M_2
< \dot M_1$ and $V_0 < V_1 < V_2$.  In these simulations only the
outburst wind is considered to be aspherical. We examine this case to
explore the effect of the post outburst wind on the nebular
morphology.  We note that the velocity difference between pre and post
outburst winds may not be physically realized. As we shall discuss in
a section 3.3 this
assumption should not effect our results significantly. The initial
conditions for each of our 9 simulations including the
index $n$ of the density function (\ie Eq.~\ref{case1b}) are given
in table 1.  

The mass loss rates use in our simulations range from $\dot M =
10^{-6}$ \mdot to $\dot M = 10^{-4}$ \mdot. Our wind velocities vary
between $100$ \kms and $1400$ \kms.  These values are representative of
what is observed in LBVs (\cite{Leitherer97}) with the highest
velocities seen in $\eta$ Car (\cite{Ebbets97}).  We note that in
giant outbursts the mass loss rates may increase by an order of
magnitude or more over what is used in this study (\cite{HumDav94}).
We have chosen not to use larger values of the mass loss rate
based on the short cooling timescales which
occur in high density winds.  Short timescales would force us to use
lower resolution simulations given constrains on computational time.
We decided that achieving higher
resolution ($256 \times 256$) was a more important goal since, based on our
results, we will be able to anticipate the enhanced cooling effects of
higher mass loss rates.  We note that the simulations were all run
until the outer shock ran off the grid.  Thus the timescale for each
simulation is essentially the dynamical age of the bubble for $R
\approx 5\times10^{17} ~cm$.

\section{Results}

\subsection{2 Wind Models with Equal $\dot M$}

Fig.~2 shows $\rho$, T, $\vec V$, and $P$ for run A after $t =
800 ~yrs$ of evolution.  Run A has an outburst stellar wind
 whose momentum flux $\dot \Pi$
is constant across the face of the star.  A number of features of the
simulation are noteworthy but it is first worthwhile to identify the
main features of the wind blown bubble.  The density map shows a bright
rim which corresponds to the shell of swept-up pre-outburst wind
material.  It is important to distinguish the composition of the material
in the shell (\ie which wind it originates
from) since its high density will make it most luminous and,
therefore, the defining feature in a real nebula.  Fig.~2 shows that the
shell develops significant asphericity due to the aspherical
driving force of the outburst wind.  For later comparison we introduce
a quantity we call the ellipticity defined as the ratio of the
distance from the star to the outer shock in the equator and 
the pole, $e = R_e/R_p$.
$e \approx
0.68$ in this simulation.  Notice also that the bubble has developed a
``waist'' which is the observational signature of a bipolar rather than
an elliptical configuration.  The density map for Run A is also shown
in Fig.~4 with the appropriate reflections to create a slice of the full
bubble.

Detailed inspection of the $\rho$ and T maps show that the shell is
thin and cold ($T \approx 10^4 K$) at all latitudes.  The narrowness
of the shell is due to the enhanced
compression which accompanies radiative losses behind the outer shock.
The role of cooling is important and is determined by the relation
between the age of the bubble and the cooling timescale. Given a shock
speed $V_s$ and a preshock density of $\rho_{pre}$ the cooling timescale
can be defined as $t_c = C {V_s}^3/\rho_{pre}$ where $C = 6 \times
10^{-35} ~g ~cm^{-6} ~s^{4}$ (\cite{Kahn76}).  The outer shock speed 
in this simulation
is on the order of 200 km/s.  Taking $\rho_{pre}$ from the mass
conservation in the pre-outburst wind gives $t_c = 150 y$ which is short
compared to the age of the bubble in Fig.~2.

The situation is quite different behind the ``reverse ''or ``inner'' shock.
At low latitudes the speed of the outburst wind is relatively small (150 km/s)
and the density is relatively high producing a short cooling time.
The reverse situation occurs at the poles where the high wind velocity
(750 km/s)
and low pre-shock density yield cooling times which are of the same order
or longer than the age of the bubble.  This produces a cap of
high temperature gas at the poles.  High thermal pressure enhances
the preferential expansion at the poles which leads to a bipolar bubble.
We note however that even when the cooling at the poles is strong enough
to drain away all shock thermalized energy the bubble which is produced
is still bipolar.

The density maps for run B and C are presented in Fig.~4.  Run
B corresponds to the case where the energy input in the wind,
$\dot E_1$, is constant across the face of the star.  The morphology
of the bubble in run B is similar to that in run A.  The principle difference
being the aspect ratio of the bubble. This can be attributed to the differing
latitudinal density profiles for the two runs.  From Eq.~(\ref{case1b})
the $\dot \Pi_1 = Const$ simulation (run A) has
$\rho \propto f(\theta)^{-2}$. The $\dot E_1 = Const$ simulation (run B) yields
$\rho \propto f(\theta)^{-3}$.  For run C the density in the wind is constant and
is set by the value in the equator where $V_1(\theta)$ has its minimum.
Since $\rho \propto \dot M/V$ via the continuity equation the outburst wind
density in run C is almost as high as that in the pre-outburst wind.  This 
produces a strong ram pressure flux ratio between the two flows.  The initial
shape of the outburst wind is strongly imprinted on the developing bubble.
Thus run C produces most bipolar shell of all three simulations 
examined in this first set of experiments.

The results presented above show that bipolar bubbles can 
form when driven by an aspherical fast wind. The differences between the 
three models also demonstrates
that the degree of bipolarity does depend on relative densities between the
aspherical fast (outburst) wind and the spherically symmetric slow pre-outburst
wind.

\subsection{2 Wind Models with $\dot M_1 > \dot M_0$}

As \cite{Langerea94} have demonstrated with stellar evolution codes, the LBV
outburst phase is likely to involve an increase in mass loss over the
pre-outburst wind.  Thus we have the case of a ``heavy'' wind expanding
into a light one.  In this situation the pre-existing circumstellar
material can not constrain the outflow.  The wind will not be
significantly decelerated by the environment until the swept-up mass is
comparable to the mass ejected by the wind.  Given that the
outburst mass loss rate can be an order of magnitude or more greater
than the pre-outburst value, the outburst wind can expand almost
ballistically for some time.  If, in addition, the outburst wind is
aspherical, then the bubble which is created will retain much of the
asphericity imprinted on the wind at the stellar surface.  In this
scenario we expect that one gets out something similar to what is put
in via the initial density and velocity distributions for the outburst wind.
To test this expectation we have run a second set of
experiments consisting of three simulations.  Runs D, E and F have
$\dot \Pi_1$, $\dot E_1$ and $\rho_1$ held constant respectively just as
in the simulations described in the last section.  In these 
simulations however $\dot M_1 = 100 \dot M_0$.   Note
that for the sake of consistency with the previous experiments
we have used the same velocity in the 
pre-outburst wind as was used in runs A through C.  
The results of \cite{Langerea94} show that the velocity
in the pre-outburst wind should be higher than that in the outburst wind.  
We do not believe such a difference will produce a profound change
in our results.  A high velocity pre-outburst wind will interact with the
ISM producing shocked gas which
backflows to the boundary of the pre-outburst wind.  Such a high pressure
region will still yield an isotropic force opposing the expansion
of the outburst wind.  Thus while the details of the flow may change, the
global properties most likely will not be affected.

Fig.~3. shows a snapshot of the evolution of run D after $t = 240yrs$.
As expected the bubble formed appears strongly bipolar with an
ellipticity of $e = 0.4$.  While such a result was expected from 
intuitive arguments
a ballistic expansion model does not tell the whole story.  Note first that
run D has reached the same size ($R_e = 5.1 \times 10^{17} $ cm) as run A
in about a third of the time giving an expansion velocity along the
pole of $V_s \approx 660$ km/s.  This is less than the outburst wind
velocity at the poles $V_1 = 750$ km/s.  Ambient material which passes through
shock can be expected to reach high temperatures of the order of $T\sim
10^7$ K where cooling is relatively ineffective.  The temperature map
clearly shows high $T$ gas at the top of the bubble.  Consideration of
the passive tracer shows that this is ambient material which has been
accelerated.  The temperature map also shows lower temperature gas at
smaller radii.  This is outburst wind material which has
passed through the inner shock which can be seen at
$R_e = 4.2 \times 10^{17} $ cm.  Consideration of the position of the
shock compared with the position expected for ballistic winds yields a
inner shock velocity of $V_s \approx 200$ \kms, consistent with the
temperature in this region (measured to $T \approx 10^6$ K). The
presence of this shock indicates that the outburst wind gas parcels are
being decelerated by the swept-up ambient material.
This deceleration occurs because of the low outburst wind
density along the poles as can see from the density map.
Recall that we are using equator to
pole velocity contrast of $q_v = 1/5$.  This produces a equator to
pole density contrast of $q_\rho = 1/{q_v}^2 = 25$.  

Runs E and F show similar morphologies though the $\rho = Const$ case 
again produces the most bipolar
configuration. We note that for Run E and F $e = 0.42$ and $0.36$
respectively.  In Fig.~4 we present $\log_{10}~\rho$ maps for runs A
through F.  In each map we have reflected the computational space about
the symmetry axis ($r = 0$) and symmetry plane ($ z = 0$) to show the
full bubble cross section and facilitate comparison.  In spite of the
differences between the runs, Fig.~4 demonstrates the principle
conclusion of our first two sets of experiments: an {\it aspherical
stellar wind can drive an aspherical bubble}.

\subsection{3 Wind Models}

In this section we present the results of our final set of numerical
experiments where we take a step closer to reality by
invoking three wind models.  In runs G, H, and I we have performed
simulations with conditions in the pre-outburst
 and outburst winds similar to run D.  In these runs however
 the outburst wind 
lasts only 30 years after which a ``post-outburst'' wind is driven into the
grid.  The characteristics of the post-outburst wind were meant to mimic
the conditions currently observed in $\eta$ Carinae in the sense that
we used a relatively low mass rate and high velocity for the wind, \ie
$\dot M_2 < \dot M_1$ and $V_2 > V_1$.  Recall that in the simulations of
GLM it was the fast post-outburst wind which was responsible for driving
the bipolar $\eta$ Car bubble.  In these simulations we are interested
in what effect this post-outburst wind will have on a bipolar bubble created
by a previously ejected dense aspherical wind.  We note that the 
post-outburst wind in all of these simulations is spherical and will 
produce an isotropic driving force.

We have run three simulations each of which utilized the $\dot \Pi =
Const$ formalism.  In these simulations it was $q_v$, the equator to
pole velocity contrast in the outburst wind, which was varied.  The
purpose of this strategy was to understand how the additional driving
force provided by the post-outburst wind would effect bubbles with different
degrees of bipolarity. 

Fig.~~5 shows a snapshot of the evolution of run H after $t = 250yrs$.
This simulation has an outburst wind with $q_v = 1/7$ and $q_{\rho} = 49$.
Unlike the last set of experiments
here the gas ejected during the outburst in these models occupies a
thin shell. The reasons for this are twofold.  First, the outburst had a
short intrinsic lifetime (which we denote as $t_1$) so its intrinsic width 
is $\delta R = V_1
t_1$. Second, the outburst material has been decelerated and compressed
from the outside by the pre-outburst wind and compressed and accelerated
from the inside by the action of succeeding post-outburst wind. Note that
the post-outburst wind has itself been decelerated through the action
of a strong inner shock.  The temperature behind this
shock at the poles is quite high ($T \approx 2. \times 10^7$ K)
reflecting the high initial velocity of the post-outburst wind.
The inner shock has a highly aspherical configuration. The
radially streaming post-outburst wind encounters this shock at oblique
angles and is refracted towards the symmetry axis (\cite{FM96}). This
produces strong shearing flows in the region behind the inner shock
which appear to give rise to instabilities in the interface
separating the shocked post-outburst and outburst flows.

In Fig.~6 we present $\log_{10} ~\rho$ maps for all three simulations
from this set of experiments.  Again it is clear that strong bipolar
morphologies develop without the need for a slow-moving disk.
Comparison between the simulations shows that decreasing $q_v$ produces
stronger bipolar morphologies.   It is
worth noting that if the expansion of the bubble were ballistic then we
would expect $e \approx q_v$ since $R_e = V_{1e} t$ and $R_p = V_{1p}
t$.  This is not the case.  Thus these bubbles show 
the result of significant hydrodynamic
shaping.  Some part of the change in the shape results from the
deceleration of the outburst wind via the previously ejected
material.  The post-outburst wind however also contributes by
accelerating the outburst wind material.  As the bubble evolves
this acceleration will have its greatest effect near the equator where
the outburst wind has been most strongly decelerated.  Thus the action
of the post-outburst wind will be to drive the bubble towards a more
spherical configuration as system evolves.  We note however that the 
post-outburst might also be aspherical.  As was noted above 
\cite{Leithererea94} have observed two components in the wind of AG Car.
An strong aspherical post outburst wind would then increase the 
elongation of the bipolar lobes. Further observational study of the
characteristics of LBV winds are warranted to determine the latitudinal 
density and velocity variations.

\section{Discussion and Conclusions} 

The results of our simulations
demonstrate that bipolar wind blown bubbles can result purely from the
action of an aspherical fast wind.  In previous studies of LBV bubbles
(FBD, NLCS, GLM) it has been assumed that a slow moving torus or disk
of gas was a necessary precondition 
for the development a bipolar bubble.  Our results indicate that
the properties of LBV bubbles may not require such a torus to form
either before (FBD, NLCS) or during (GLM) the outburst.

Our scenario has a number of attractive features.  First it is both
observationally and theoretically motivated.  From the observational
side there is already evidence that LBV winds can take on aspherical
velocity and density distributions.  From the theoretical side the
theory of \cite{LamPaul91} have demonstrated that "bistable" winds are
possible around massive hot stars.  In addition, the diversity of shapes
of LBV bubbles (NLCS) may be difficult to achieve with 
pre-exiting disk models.   The models presented here
can recover the diversity of shapes simply by changing $q_v$ though it
is certainly true that this also begs the question of what drives the
velocity contrast in the fast wind. 

Currently it is unclear what form the mass distribution takes in the
lobes of LBV bubbles.  One critical test of the different models for
LBV nebula shaping would entail comparison of latitudinal variations of
mass.  If the caps of a bipolar lobe have densities that are comparable
to that in the lobe's flanks it would present difficulties
for the slow torus models. In a bipolar bubble resulting from a
spherical fast wind driving into a slow torus the caps of the
bubble should be the location of the lowest density.    For aspherical
fast winds however high or equal density in the poles poses no
significant problem since we are then seeing a signature of the
latitudinal dependence of the fast wind density (consider Runs C and F
in Fig.~3). We wish to note also that \cite{Currie96b} and 
\cite{Morseea97} have found that
the shape of $\eta$ Car is best matched by by a geometry which can be described
as a "double flask" rather than an two oscullating spheres.
Based on comparison of published results
the models presented here seem to do a better job of recovering such a
shape than either FBD, NLCS or GLM.

It is noteworthy however that the scenario presented here would not
produce the equatorial ``skirt'' seen surrounding the waist of the
homunculus in $\eta$ Car.  The presence of that feature is what
motivated the original application of the GISW slow torus models.
Within the current formulation of the aspherical fast wind model there
is not a likely means of producing such a feature.  A few points are
worth noting however.  The equatorial
skirt is not a continuous or even axisymmetric feature. Thus whatever
its origin it is hard to imagine that it can be the agent which
constricts a spherical fast wind and produces the bipolar bubble.  In
addition there are a number of ``spikes'' extending beyond, but
connecting with the homunculus that have been reported to have
velocities in excess of 1000 km/s (\cite{Meaburnea96}).  They yield
dynamical timescales $\le t_o$ where $t_o$ is the time since the
outburst.  Thus it is possible that the equatorial skirt is actually a
spray of material which ejected at some point after the outburst of
1849 and which was decelerated by its passage through the dense shell
of the outburst wind. The non-axisymmetric distribution of the skirt may
then be a consequence of the intrinsic pattern of ejecta or of impulsive
instabilities which will occur when the ejecta is driven through the
outburst wind.

Regardless of the answer to this issue the results presented here show
that there are two very different scenarios for the formation
of LBV bubbles.  Either they form via the interaction of a spherical fast wind
driving into an aspherical slow wind (a slow torus) or they form from
an aspherical fast wind driving into an isotropic pre-existing environment.
This embarrassment of riches can be eventually be dealt with by comparing
the latitudinal distributions of mass and momentum observed in
real bipolar LBVs with what is predicted for the various models.  Such an
approach was used successfully by \cite{CherMas92} in evaluating different 
models of molecular outflow formation in YSOs.  This project is currently 
in progress.

\acknowledgments
We wish to thank  Jon Morse, Jon Bjorkman, Tom Jones and Stan Owocki for 
the very useful and enlightening discussions on this
topic.
Support for AF was provided
by NASA grant HS-01070.01-94A
from the Space Telescope Science Institute, which is operated by
AURA Inc under NASA contract NASA-26555.  Additional support for AF
came from NSF Grant AST-9702484 and the Minnesota Supercomputer Institute.
Partial support for DR was provided from Seoam Scholarship Foundation.

\clearpage

\begin {table}[hbta]
\caption {Initial Conditions For Runs A - H}

\begin {center}
\begin {tabular} {lllllllll} \hline
{run}  & {${\dot M}_{\rm 0}$} & {$V_{\rm 0}$} & {${\dot M}_{\rm 1}$} 
& {${V}_{\rm 1}$} &  {${\dot M}_{\rm 2}$} 
& {${V}_{\rm 2}$} &   {$q_v$} & {Con. Index}\\
\hline
A  &  $1 \times 10^{-4}$ & 100  & $1 \times 10^{-4}$ & 150 & 
NA & NA & $0.2$ & -2\\

B  &  $1 \times 10^{-4}$ & 100  & $1 \times 10^{-4}$ & 150 & 
NA & NA & $0.2$ & -3\\

C  &  $1 \times 10^{-4}$ & 100  & $1 \times 10^{-4}$ & 150 & 
NA & NA & $0.2$ & 0\\

D  &  $1 \times 10^{-6}$ & 100  & $1 \times 10^{-4}$ & 150 & 
NA & NA & $0.2$ & -2\\

E  &  $1 \times 10^{-6}$ & 100  & $1 \times 10^{-4}$ & 150 & 
NA & NA & $0.2$ & -3\\

F  &  $1 \times 10^{-6}$ & 100  & $1 \times 10^{-4}$ & 150 & 
NA & NA & $0.2$ & 0\\

G  &  $1 \times 10^{-6}$ & 100  & $1 \times 10^{-4}$ & 150 & 
$1 \times 10^{-6}$ & 1400 & $0.3$ & -2\\

H  &  $1 \times 10^{-6}$ & 100  & $1 \times 10^{-4}$ & 150 & 
$1 \times 10^{-6}$ & 1400 & $0.14$ & -2\\

I  &  $1 \times 10^{-6}$ & 100  & $1 \times 10^{-4}$ & 150 & 
$1 \times 10^{-6}$ & 1400 & $0.1$ & -2\\

\end {tabular}
\end {center}
\end {table}

\clearpage

\begin{center}
{\bf FIGURE CAPTIONS}
\end{center}
\begin{description}

\item[Fig.~1]{Initial conditions for aspherical fast stellar wind.
The variation of stellar wind velocity and density with
latitude are shown for a model with A = 0.1 and B = 2. Note the equator
is at $\theta = 90^o$. The velocity and
density at the
equator are set to 1.  The density plot shows the three different cases
considered in text. Solid line: momentum input $\dot \Pi = const$ across
face of star. Dashed line: energy input $\dot E = const$. Dashed-dot line:
density $\rho = const$.

\item[Fig.~2]{Run A after $800 yrs$. Shown are greyscale maps of
$\log_{10}$ number density, temperature, velocity (with vectors
superposed), and pressure.  Note that dark scales correspond
to high values of the variables.  The range of variables are as follows:
$.5 < \log_{10}( n/cm^{-3})  <  3.7$, 
 ~$3.9 < \log_{10}(T /K)   <  6.8$, 
 ~$6.1  <  \log_{10} (V/ cm s^{-1}) <  7 .8$, and 
$-11.3  <   \log_{10} (P /dynes ~cm^{-2})  < -7.8$}

\item[Fig.~3]{Run D after $240 yrs$. The maps and parameter ranges are
the same as in Fig.~2}

\item[Fig.~4]{$\log_{10}$ grayscale density maps of
runs A through F.  Top Left: Run A. Top Center: Run B. Top Right: Run C.
Bottom Left: Run D. Bottom  Center: Run E. Bottom  Right: Run F. 
Note that here that dark scales correspond to low values of the variables.
Runs A and B are shown at $t \sim 800 yrs$. Runs
C through F are shown at $t \sim 250 yrs$. 
 Each image is $1.2x10^{18}$ cm square.}

\item[Fig.~5]{Run G after $250 yrs$. Shown are greyscale maps of
$\log_{10}$ number density, temperature, velocity (with vectors
superposed), and pressure.  Note that dark scales correspond
to high values of the variables.  The range of variables are as follows:
$.5 < \log_{10}( n/cm^{-3})  <  3.1$, 
 ~$3.9 < \log_{10}(T /K)   <  7.5$, 
 ~$6.5  <  \log_{10} (V/ cm ~s^{-1})  <  8.1$, and 
$-11.3  <   \log_{10} (P /dynes ~cm^{-2}) < -7.7$}
 
\item[Fig.~6]{$\log_{10}$ grayscale density maps of
runs G through I.  Left: Run G. Center: Run H. Right: Run I. 
Note that here that dark scales correspond to low values of the variables.}
Each image is $1.2x10^{18}$ cm square.}
\end{description}

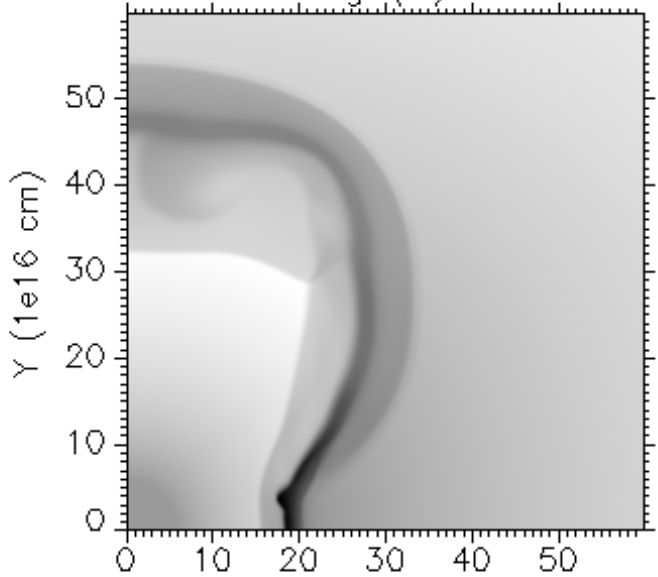
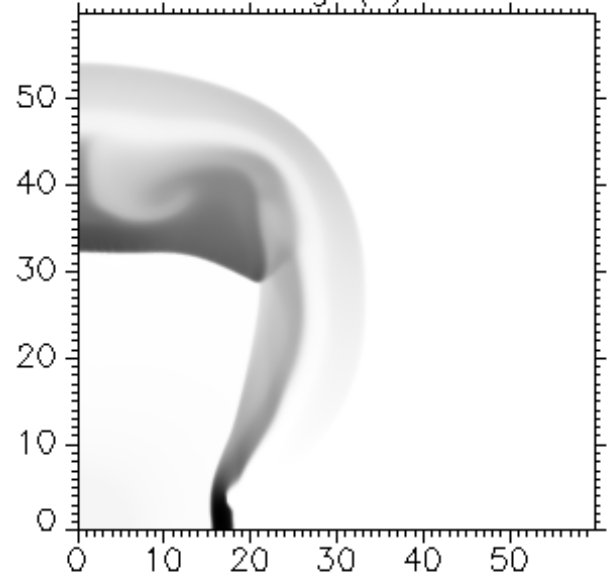
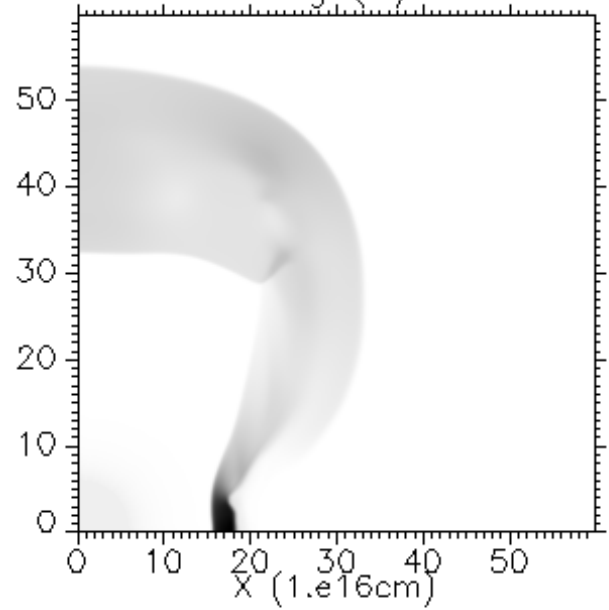
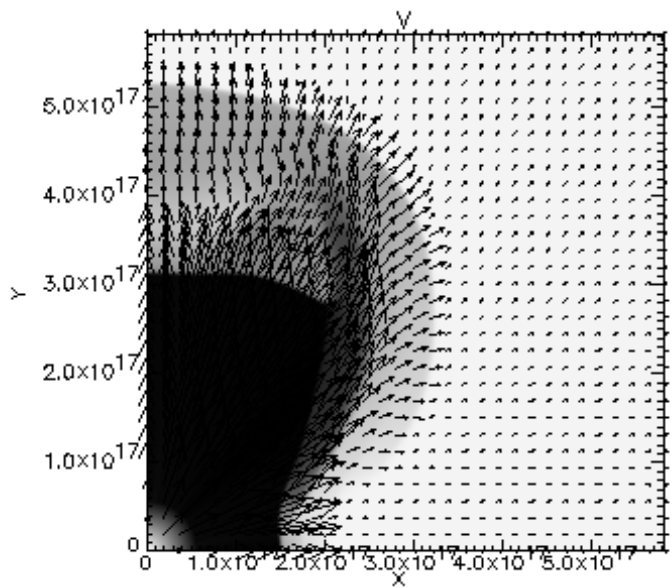
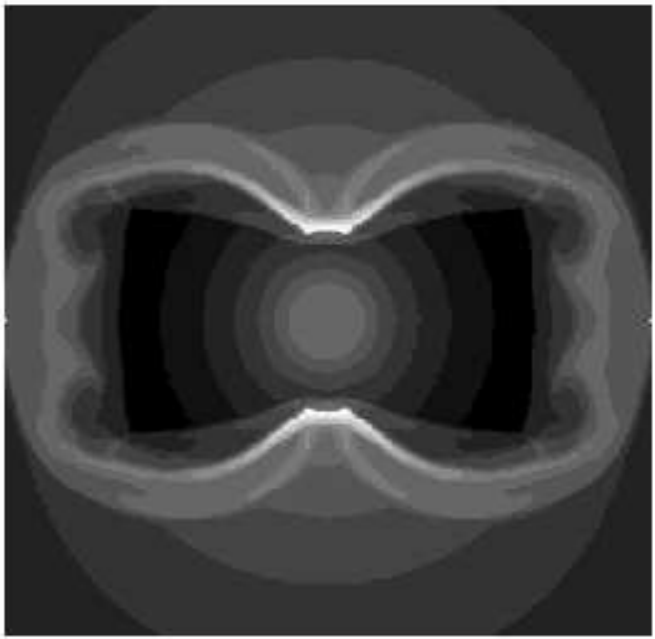
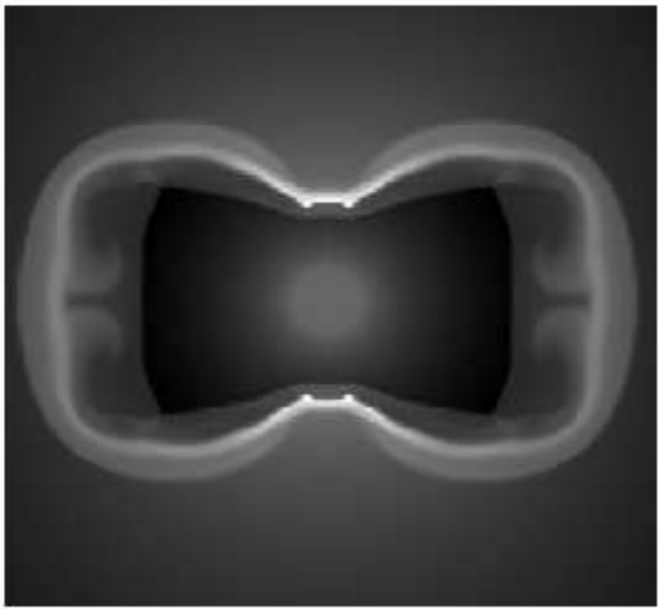
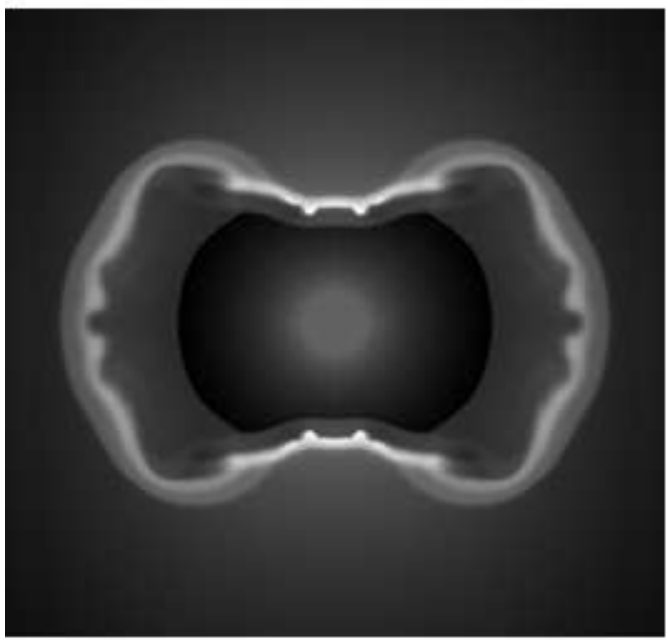
\end{document}